\documentclass[twocolumn,floats,floatfix,superscriptaddress,aps,pra]{revtex4-1}
\usepackage{subfigure}
\usepackage{psfrag,graphicx}
\usepackage{bm,color}
\usepackage{latexsym}
\usepackage{epstopdf}
\usepackage[english]{babel}
\usepackage{amsmath,amssymb,amsfonts}
\usepackage{appendix}
\usepackage{bbold}

\definecolor{mygrey}{gray}{0.35}
\definecolor{myblue}{rgb}{0.2,0.2,0.8}
\definecolor{myzard}{cmyk}{0,0,0.05,0}
\definecolor{mywhite}{rgb}{1,1,1}
\definecolor{mywhite}{rgb}{1,1,1}
\definecolor{myred}{rgb}{1,0.,0.3}

\usepackage[colorlinks=true,citecolor=myblue,linkcolor=myred]{hyperref}

\def\be{\begin{equation}}
\def\ee{\end{equation}}
\def\bse{\begin{subequations}}
\def\ese{\end{subequations}}
\def\ba{\begin{align}}
\def\enda{\end{align}}
\def\bi{\begin{itemize}}
\def\ei{\end{itemize}}

\def\bt{\begin{tabular}}
\def\et{\end{tabular}}

 \newcommand{\ket}[1]{|#1\rangle}

\def\ket#1{\vert #1 \rangle}

\def\Black{}

\def\from{\leftarrow}
\def\to{\rightarrow}
\def\adbphase{\eta}

\def\i{\textrm{i}}
\def\f{\textrm{f}}

\def\U{\mathbf{U}}
\def\I{\mathbf{I}}
\def\H{\mathbf{H}}
\def\F{\mathbf{F}}
\def\F{\bm{\Phi}}
\def\Re{\text{Re}}
\def\Im{\text{Im}}

\begin{document}

\author{Kaloyan N. Zlatanov}
\affiliation{Department of Physics, St Kliment Ohridski University of Sofia, 5 James Bourchier blvd, 1164 Sofia, Bulgaria}
\affiliation{Institute of Solid State Physics, Bulgarian Academy of Sciences, Tsarigradsko chaussée 72, 1784 Sofia, Bulgaria}
\author{Nikolay V. Vitanov}
\affiliation{Department of Physics, St Kliment Ohridski University of Sofia, 5 James Bourchier blvd, 1164 Sofia, Bulgaria}

\title{Generation of arbitrary qubit states by adiabatic evolution split by a phase jump}

\date{\today }

\begin{abstract}
We propose a technique for accurate, flexible and robust generation of arbitrary pre-selected coherent superpositions of two quantum states.
It uses a sequence of two adiabatic pulses split by a phase jump serving as a control parameter.
Each pulse has a chirped detuning, which induces a half crossing, and acts approximately as a half-$\pi$ pulse in the adiabatic regime.
The phase jump is imprinted onto the population ratio of the created superposition state.
Of the various possible relations between the two pulses, we select the case when the Rabi frequency and the detuning of the second pulse are mirror images of those of the first pulse, and the two detunings have opposite signs.
Then the mixing angle of the created superposition state depends on the phase jump only.
For other arrangements, the superposition mixing angle is shifted by the dynamic phases of the propagators, which makes these cases suitable for state tomography.
This sandwich setup comes along with the advantage  that it reduces the error $\epsilon$ of each individual pulse down to $4\epsilon^2$ overall.
Therefore, the proposed technique combines the benefits of \emph{robustness} stemming from adiabatic evolution with \emph{accuracy} generated by the twin-pulse error suppression, and \emph{flexibility} of the created superposition state controlled by the value of the phase jump $\phi$.
In addition to the general analysis, we present a simple exactly soluble trigonometric model in order to illustrate the proposed technique.
In this model, when the pulse area $A$ increases the nonadiabatic oscillations are damped as $A^{-1}$ for a single pulse and $A^{-2}$ for the two-pulse sequence.
Finally, the proposed technique is iteratively extended to sequences of $N=2^n$ pulses by concatenating half-$\pi$ sequences and splitting them by a phase jump, thereby further reducing the nonadiabatic error $\epsilon$ to $(2\epsilon)^{N}$.
This makes the proposed technique suitable for generating high-fidelity quantum rotation gates even when starting with errant pulses.
\end{abstract}

\maketitle

\section{Introduction}

Coherent superposition states are one of the cornerstones of contemporary quantum physics.
They are essential in a variety of quantum phenomena, such as dark resonances \cite{Alzetta1976,Arimondo1976,Alzetta1979}, electromagnetically induced transparency \cite{Boller1991,Harris1997,Fleischhauer2005}, light amplification without inversion \cite{Kocharovskaya1992}, photon memories \cite{Lukin2003}, conversion efficiency improvements in high harmonic generation \cite{Watson1996}, and nonlinear optics \cite{Jain1996}, to mention just a few.
Coherent superpositions of quantum states are crucial in quantum information and quantum technologies in general \cite{Nielsen2000}.
For example, the Hadamard gate which, starting from a single qubit state, creates a maximally-coherent equal superposition of two states, is a basic quantum gate at the core of most quantum protocols.

Due to their numerous applications a number of techniques have been developed for their generation.
The simplest technique is a direct linkage between the two states of the superposition with a resonant  pulse with a temporal area of $\pi/2$ \cite{Allen1975,Shore1990}.
This technique, however, is not robust to experimental errors since the superposition states are very sensitive to variations in the experimental parameters, including the amplitude, the duration, and the detuning of the field.
For that matter any alternative technique is required most of all to be robust, and, if possible, technically undemanding.

Adiabatic techniques are a viable alternative to resonant pulses of precise temporal area.
They offer robustness to variations in various experimental parameters at the expense of larger pulse area and hence longer interaction duration \cite{Vitanov2001}.
In two-state systems, two basic regimes of adiabatic evolution are distinguished.
Complete population transfer (CPT) occurs when the energies of the two states cross at a certain instant of time \cite{Landau1932,Zener1932,Stuckelberg1932,Majorana1932}.
On the contrary, complete population return (CPR) takes place when these energies do not cross \cite{Vitanov2001} and it has interesting applications too \cite{Haroche2013}.

In three-state systems, adiabatic evolution is used in the famous stimulated Raman adiabatic passage (STIRAP) technique \cite{Vitanov2017}.
STIRAP is the most popular tool to completely transfer the population between the two end states 1 and 3 in a three-state chain $1-2-3$, whenever the direct linkage $1 \to 3$ is not possible, e.g. due to electric-dipole selection rules.
A unique feature of STIRAP is that in the adiabatic limit the (usually lossy) middle state 2 remains unpopulated, even transiently, because the population remains in the so-called dark state, which is a coherent superposition of states 1 and 3 only.
This remarkable feature makes STIRAP largely immune to losses from state 2.
Extensions of STIRAP to more states have also been proposed and implemented \cite{Vitanov2017}.
We note that there exist other adiabatic techniques in three-state and multistate systems, which use the level crossing concept \cite{Vitanov2001, Rangelov2005, Oberst2008}.

Variations of the above adiabatic techniques have been proposed and demonstrated also for generation of coherent superposition states.
In two-state systems, adiabatic evolution has been used in a technique known as half-SCRAP \cite{Yatsenko2002} and the closely related two-state STIRAP \cite{Vitanov2006,Yamazaki2008}.
In both cases pulse shaping and chirping are designed such that their time dependences resemble the delayed-pulse ordering of conventional STIRAP.
In three-state chains, STIRAP has been modified in a configuration known as fractional STIRAP \cite{Marte1991,Weitz1994,Vitanov1999}, in which the Stokes pulse arrives before the pump pulse but the two pulses vanish simultaneously.
This leads to the creation of a coherent superposition of the two end states 1 and 3.
Tripod-STIRAP \cite{Unanyan1998,Theuer1999,Vewinger2003}, an extenstion of STIRAP wherein a single state is coupled to three other states, has also been used for the generation of coherent superpositions of these three states or two of them.
We also note a technique for creation of coherent superposition states and for navigation between them by quantum Householder reflections \cite{Ivanov2007}.


In a previous paper \cite{Zlatanov2017} we have introduced a single-pulse technique to exploit the robustness of adiabatic evolution for generation of arbitrary coherent superpositions of two states, including maximally coherent states, i.e. with transition probability of $\frac12$.
The final superposition state is determined by the initial and final ratios of the field's amplitude and its detuning.
In particular, if the detuning is chirped such that it starts from a nonzero initial value and ends up on resonance (or vice versa), in a ``half-crossing'' pattern, while the Rabi frequency changes in the opposite manner, such a pulse produces a transition probability of $\frac12$ in the adiabatic limit.
An extension of this technique to three states has been experimentally demonstrated in a trapped-ion experiment, with a fidelity close to the 99.99\% quantum computation benchmark level \cite{Randall2018}.
Such a mechanism requires a precise control over the initial and final values of the Rabi frequency $\Omega(t)$ and the detuning $\Delta(t)$.
When such a control is difficult, one may seek a different efficient control parameter, while preserving the robustness.

To this end, here we propose using a phase jump in the field amplitude as a control parameter.
Phase jumps have proved to have a dramatic influence over the evolution of the system \cite{Vitanov2007,Torosov2007}.
Furthermore, phase jumps are the key control parameter in robust coherent control techniques, such as the composite pulses \cite{Levitt1979, Freeman1980, Levitt1982, Levitt1986, Wimperis1994, Levitt2001, Vandersypen2004, Levitt2007}, which are a popular control tool for compensating systematic field errors \cite{Manu2015,Khaneja2017,Timoney2008,Monz2009,Mount2015,Zanon-Willette2018,Berg2015,Wang2012,Rong2015,Aiello2013,Low2016,Torosov2011PRL,Schraft2013,Merrill2014review,Genov2014PRL}.
Specifically, we introduce a technique for creation of arbitrary pre-selected coherent superposition states of a qubit.
It uses a combination of two adiabatic pulses, each producing a single-pulse transition probability of $p\approx \frac12$, divided by a phase jump.
In this manner we harness the robustness of the adiabatic evolution but shift the control solely to the phase jump.
%
By selecting the Rabi frequency $\Omega(t)$ and the detuning $\Delta(t)$ of the first pulse, we demonstrate in Sec.~\ref{sec:two} how the second pulse must be selected such that an error $\epsilon$ in  the transition probability $p = \frac12-\epsilon$ of each pulse can be reduced to $O(\epsilon^2)$ for the two-pulse sequence.
Then we provide  examples with a simple analytically solvable trigonometric model in Sec.~\ref{sec:three}.
Finally, we discuss the extension of this two-pulse technique to sequences of multiple pulses in Sec.~\ref{sec:multi}, which further increase the accuracy.



\section{Twin pulses split by a phase jump}\label{sec:two}

\subsection{Adiabatic solution for a single pulse}

We assume that the Hamiltonian is given in the symmetric form
\be\label{Hamiltonian}
\mathbf{H}(t)=\tfrac{1}{2} \hbar \left[\begin{array}{cc} -\Delta (t) & \Omega (t) \\   \Omega (t) & \Delta (t) \end{array}\right]  ,
\ee
where $\Omega(t)$ is the Rabi frequency of the interaction and $\Delta(t)$ is the system-field frequency offset (the detuning).
Both $\Omega(t)$ and $\Delta(t)$ are assumed real, unless a dedicated phase shift is applied to $\Omega(t)$.
The Hamiltonian can be written in terms of the Pauli matrices $\bm\sigma_k$ ($k=x,y,z$) also as
\be\label{Hamiltonian-Pauli}
\mathbf{H}(t) = \tfrac{1}{2} \hbar \Omega(t) \bm\sigma_x - \tfrac12 \hbar \Delta(t) \bm\sigma_z.
\ee

The evolution of the two-state system is governed by the Schr\"odinger equation,
\be\label{SEq}
{\rm i}\hbar\frac{d}{dt}\mathbf{c}(t)=\mathbf{H}(t)\mathbf{c}(t),
\ee
where $\mathbf{c}(t)= [ c_1(t),c_2(t)]^T$ is the state vector comprising the probability amplitudes, with some specified values $c_1(t_{\i})$ and $c_2(t_{\i})$ at the initial time $t_{\i}$.
The propagator $\mathbf{U}(t,t_{\i})$ links the initial values of the probability amplitudes to their values at any time $t$,
\be\label{evolve-U}
\mathbf{c}(t)=\mathbf{U}(t,t_{\i})\mathbf{c}(t_{\i}).
\ee
The propagator satisfies the Schr\"odinger equation \eqref{SEq},
\be\label{SEq-U}
{\rm i}\hbar\frac{d}{dt} \mathbf{U}(t,t_{\i}) = \mathbf{H}(t) \mathbf{U}(t,t_{\i}),
\ee
with the initial condition $\mathbf{U}(t_{\i},t_{\i}) = \mathbf{I}$.
Of special interest is the propagator $\mathbf{U}(t_{\f},t_{\i})$ at the end of the interaction, at time $t_{\f}$.
The propagator has the SU(2) symmetry and hence can be parameterized as
\be \label{U}
\mathbf{U}(t_{\f},t_{\i}) = \left[\begin{array}{cc} a & -b^{\ast} \\   b & a^{\ast} \end{array}\right].
\ee
where $a$ and $b$ are the complex-valued Cayley-Klein parameters, obeying the condition  $|a|^2+|b|^2=1$.

The Bloch variables relate to the propagator through the density matrix evolution, $\bm{\rho}(t) = \mathbf{U}(t,t_{\i}) \bm{\rho}(t_{\i}) \mathbf{U}(t,t_{\i})^{\dagger}$ as $u(t) = 2\Re \rho_{12}(t)$, $v(t) = 2\Im \rho_{12}(t)$, and $w(t) = \rho_{22}(t) - \rho_{11}(t)$.
For Bloch variables starting at  $u_{\i} = v_{\i} = 0, w_{\i} =-1$ (meaning that the system is initially in state 1), the relation to the propagator elements reads
\be\label{uvw(ab)}
u_{\f} = 2 \text{Re} (a^* b),\quad v_{\f} = 2 \text{Im} (a^* b), \quad w_{\f} = |b|^2-|a|^2.
\ee
%
The adiabatic solution for the Bloch vector reads \cite{Ivanov2005,Zlatanov2017}
\bse\label{Bloch adiabatic}
\begin{eqnarray}
u_{\f} &=&\frac{\Omega _{\i}\Omega _{\f}+\Delta _{\i}\Delta _{\f}\cos \adbphase }{\Lambda _{\i}\Lambda _{\f}}u_{\i}-\frac{\Delta _{\f}\sin \adbphase }{\Lambda _{\f}}v_{\i}  \notag \\
&&+\frac{\Delta _{\i}\Omega _{\f}-\Omega _{\i}\Delta _{\f}\cos \adbphase }{\Lambda
_{\i}\Lambda _{\f}}w_{\i},  \label{ad.uf} \\
v_{\f} &=&\frac{\Delta _{\i}\sin \adbphase }{\Lambda _{\i}}u_{\i}+\cos \adbphase v_{\i}-\frac{\Omega _{\i}\sin \adbphase }{\Lambda _{\i}}w_{\i},  \label{ad.vf} \\
w_{\f} &=&\frac{\Omega _{\i}\Delta _{\f}-\Delta _{\i}\Omega _{\f}\cos \adbphase }{\Lambda _{\i}\Lambda _{\f}}u_{\i}+\frac{\Omega _{\f}\sin \adbphase }{\Lambda _{\f}} v_{\i}  \notag \\
&&+\frac{\Delta _{\i}\Delta _{\f}+\Omega _{\i}\Omega _{\f}\cos \adbphase }{\Lambda_{\i}\Lambda _{\f}}w_{\i},  \label{ad.wf}
\end{eqnarray}
\ese
where the subscripts $\i$ and $\f$ refer to the values of the respective variables at the initial and final times $t_{\i}$ and $t_{\f}$.
Here
\bse
\begin{align}
\Lambda (t) &=\sqrt{\Omega(t)^2 +\Delta(t)^2},  \label{omega} \\
\adbphase &=\int_{t_{\i}}^{t_{\f}}\Lambda (t)dt.  \label{phase}
\end{align}
\ese
This adiabatic solution applies to pure and mixed states as well, provided the adiabatic evolution is completely coherent.
The condition for adiabatic evolution is \cite{Vitanov2001}
\be\label{adiabatic condition}
\Lambda (t)\gg |\dot{\vartheta}(t)|
= \frac{\left|\Delta(t) \partial_t\Omega(t) - \Omega(t)\partial_t\Delta(t)\right|} {\Lambda^2}
\ee
with $\vartheta (t) = \tan ^{-1}[\Omega (t)/\Delta (t)]$.

If the system is initially in state $\ket{1}$, then $u_{\i}=v_{\i}=0$, $w_{\i}=-1$, and the adiabatic solution in the end reads
\bse\label{adiabatic solution-1}
\begin{align}
u_{\f} &= -\frac{\Delta _{\i}\Omega _{\f}-\Omega _{\i}\Delta _{\f}\cos \adbphase }{\Lambda _{\i}\Lambda _{\f}},  \label{u.coherent} \\
v_{\f} &= \frac{\Omega _{\i}\sin \adbphase }{\Lambda _{\i}}, \label{v.coherent} \\
w_{\f} &= -\frac{\Delta _{\i}\Delta _{\f}+\Omega _{\i}\Omega _{\f}\cos \adbphase }{\Lambda _{\i}\Lambda _{\f}}.  \label{w.coherent}
\end{align}
\ese
Therefore the single-pulse transition probability $p = (w_{\f}+1)/2$ is
\be\label{adiabaticprob}
p= \frac{1}{2} -\frac{\Delta_{\i}\Delta_{\f}}{2\Lambda_{\i}\Lambda_{\f}} -\frac{\Omega_{\i}\Omega_{\f}}{2\Lambda_{\i}\Lambda_{\f}}\cos2\eta.
\ee

\begin{figure}[tb]
\bt{cc}
\includegraphics[width=0.45\columnwidth]{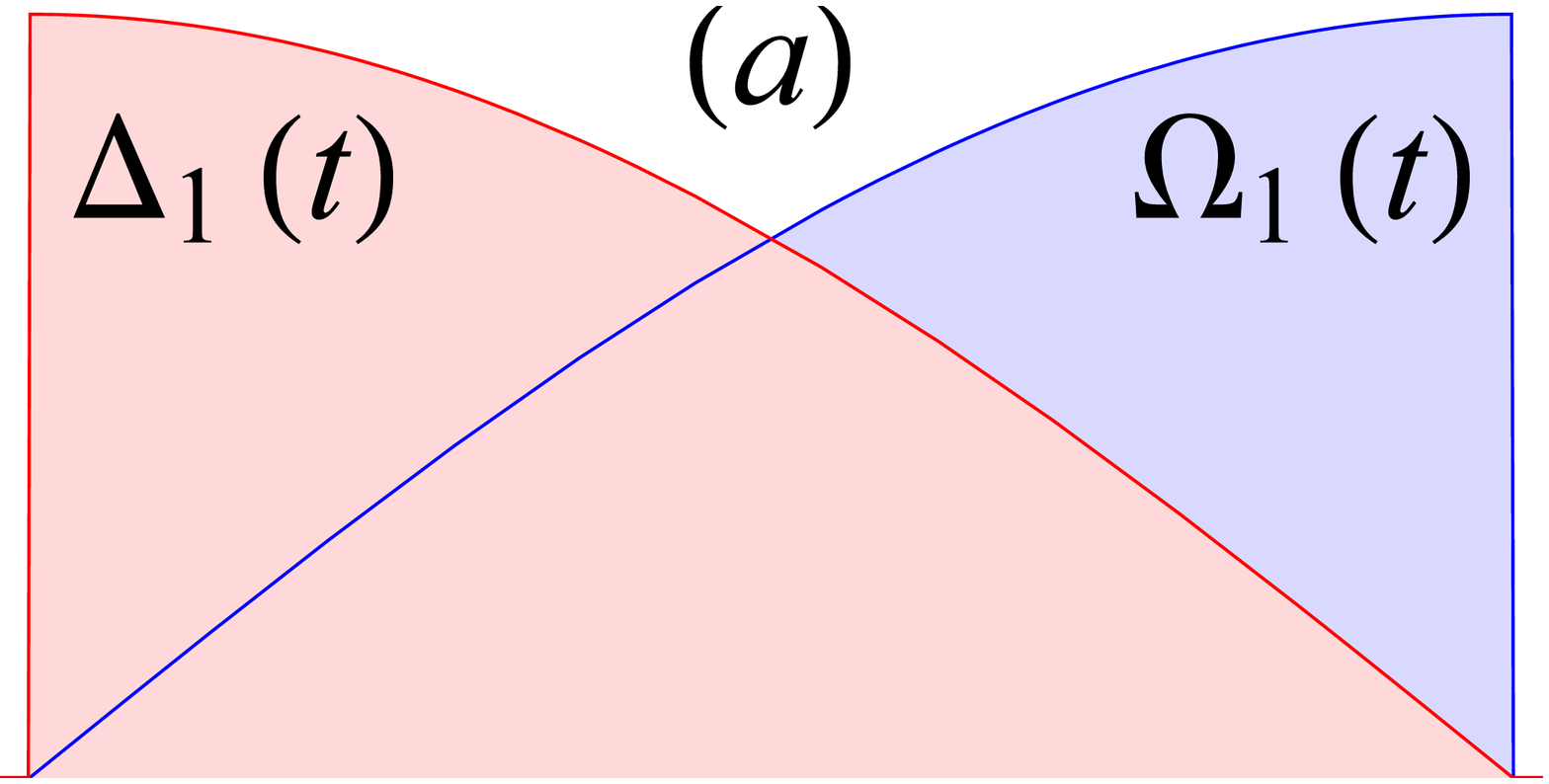} & \includegraphics[width=0.45\columnwidth]{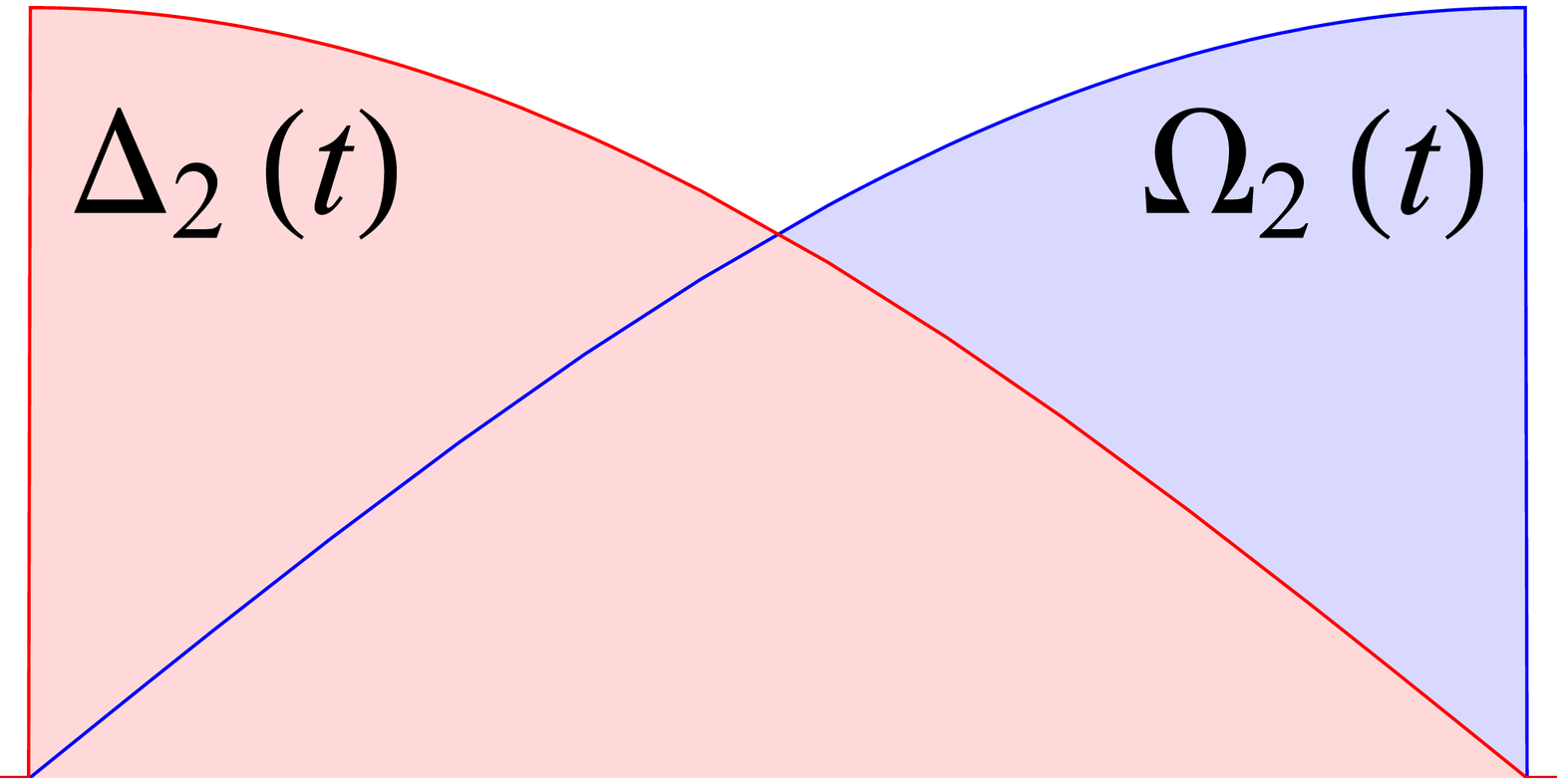} \\
\includegraphics[width=0.45\columnwidth]{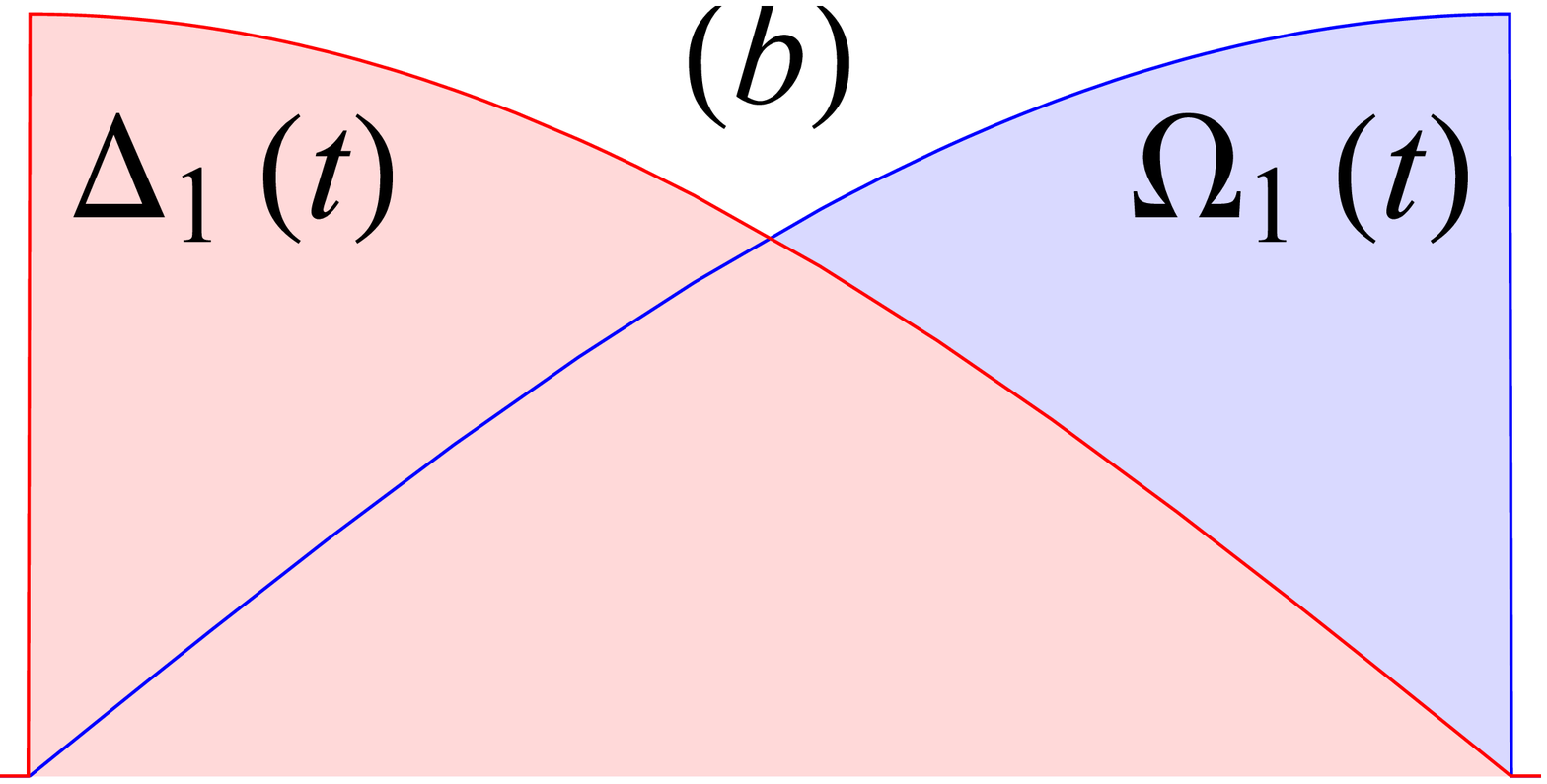} & \includegraphics[width=0.45\columnwidth]{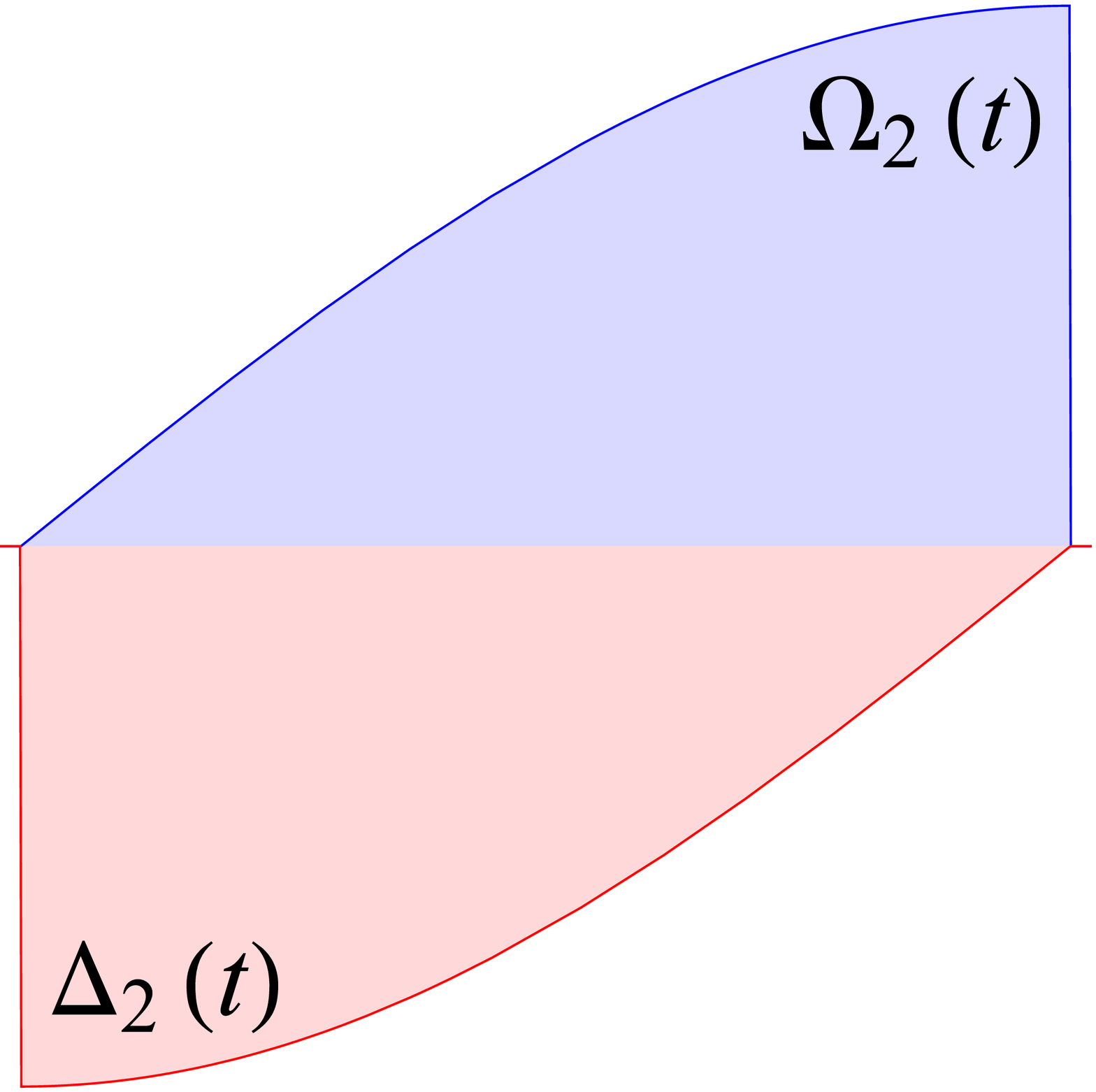} \\
\includegraphics[width=0.45\columnwidth]{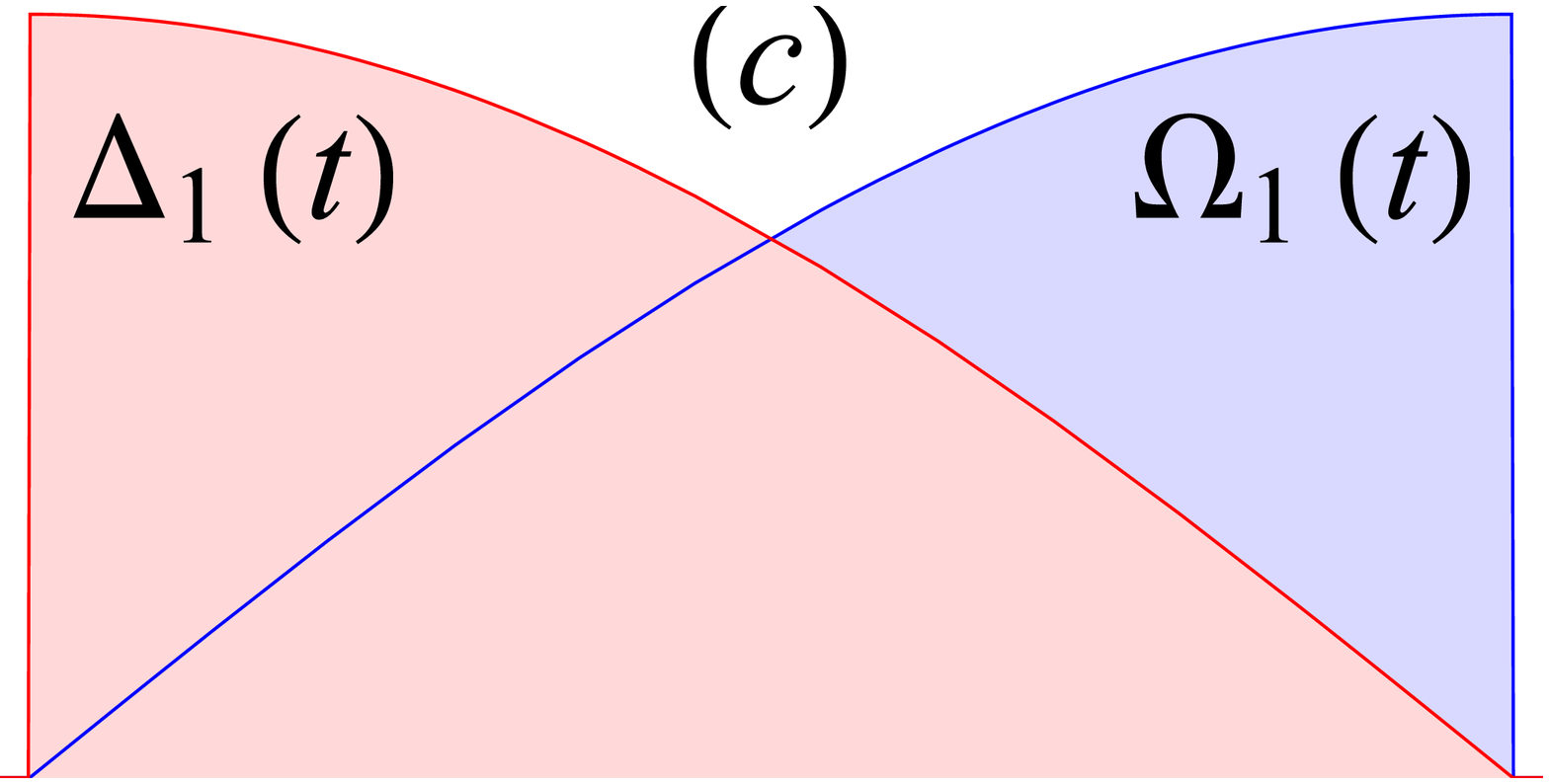} & \includegraphics[width=0.45\columnwidth]{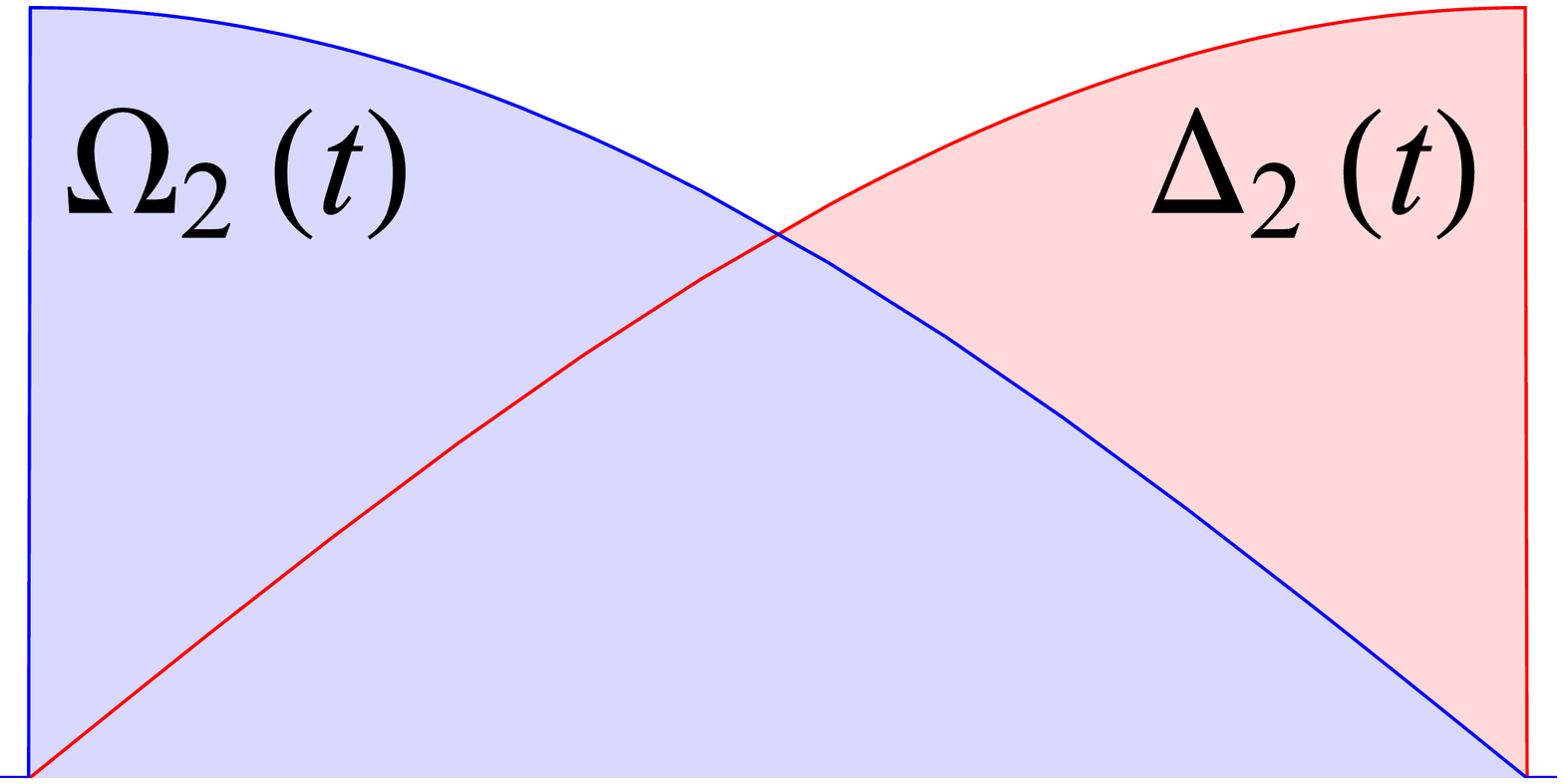} \\
\includegraphics[width=0.45\columnwidth]{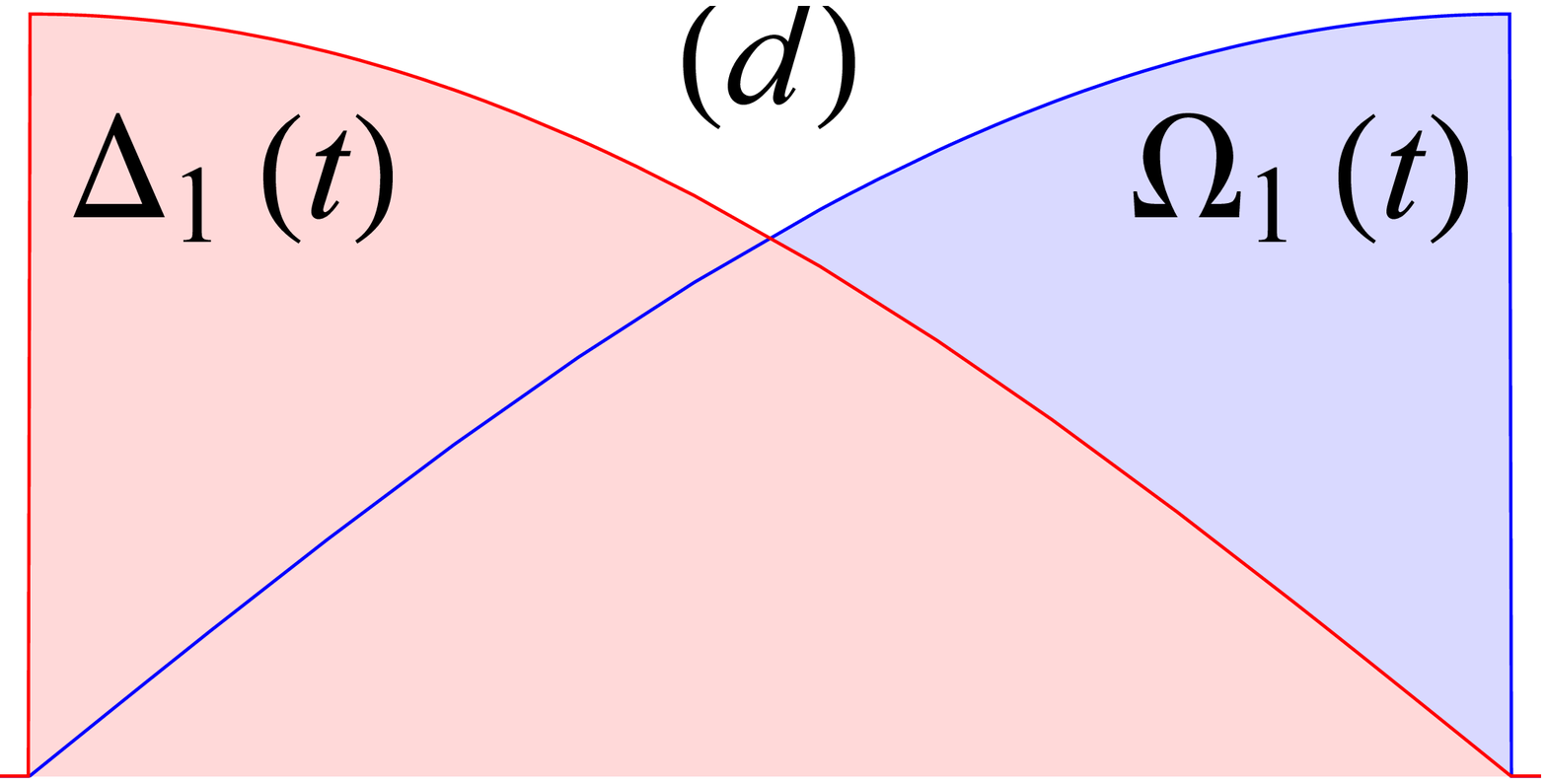} & \includegraphics[width=0.45\columnwidth]{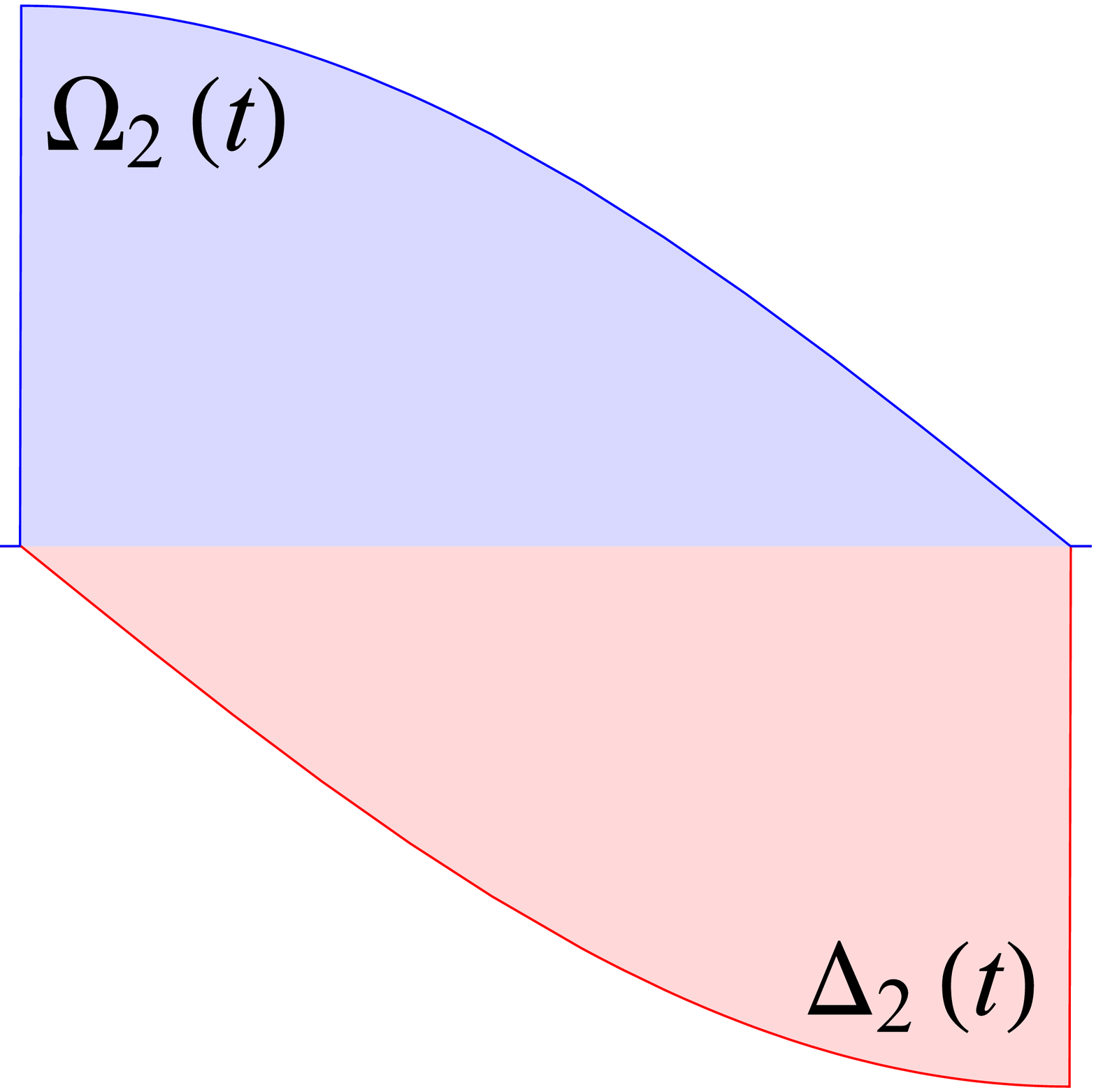}
\et
\caption{(Color online)
Various choice of pulse pairs for the double-pulse scenario.
The first pulse (on the left) is the same in all cases, with the Rabi frequency $\Omega(t)$ and the detuning $\Delta(t)$ changing in opposite directions, from a zero to a nonzero value and vice versa.
The second pulse (on the right) contains a phase jump of $\phi$ (not shown) in the Rabi frequency $\Omega(t)$.
The different cases are:
(a) The second pulse is identical to the first one.
(b) The Rabi frequency $\Omega(t)$ of the second pulse is the same as the first pulse but the detuning $\Delta(t)$ has the opposite sign.
(c) The Rabi frequency $\Omega(t)$ and the detuning $\Delta(t)$ of the second pulse are mirror images to the ones of the first pulse.
(d) The same as (c) but the detuning changes sign.
}\label{fig:shapes}
\end{figure}

In the adiabatic regime, the Rabi frequency's and detuning's initial values $\Omega_{\i},\Delta_{\i}$ and final values $\Omega_{\f},\Delta_{\f}$ determine the final position of the Bloch vector.
Therefore, an appropriate choice of these values can give any desired state on the Bloch sphere.
%
Using this leeway, recently \cite{Zlatanov2017,Randall2018} we showed how adiabatic evolution can be used to create arbitrary pre-selected coherent superposition states.
In particular, if
\bse
\be\label{case 1}
0 \stackrel{t_{\i}\from t}{\longleftarrow } \frac{\Omega (t)}{\Delta (t)} \stackrel{t \to t_{\f}}{\longrightarrow }  \infty,
\ee
or
\be \label{case 2}
\infty \stackrel{t_{\i}\from t}{\longleftarrow } \frac{\Omega (t)}{\Delta (t)} \stackrel{t \to t_{\f}}{\longrightarrow }  0,
\ee
\ese
then in each case a maximally coherent superposition ($p=\frac12$) of states 1 and 2 is created when starting from state 1.
For example, these two cases are implemented by the combinations of the Rabi frequency $\Omega(t)$ and the detuning $\Delta(t)$ shown in Fig.~\ref{fig:shapes}.
If one (or both) of the asymptotic values of the ratio $\Omega (t)/\Delta (t)$ is different from $0$ or $\infty$, then an unequal coherent superposition state is created.

In order to create a coherent superposition state with this method, two conditions must be fulfilled.
First, the adiabatic condition \eqref{adiabatic condition} must be satisfied; away from the adiabatic limit the efficiency of this process may drop considerably.
Second, the initial and final values $\Omega_{\i,\f}$ and $\Delta_{\i,\f}$ must be well controlled; small deviations from these values reduce the fidelity of the target state.

Here we show that by taking two such imperfect pulses, and phase shift the second one with respect to the first, we can considerably reduce the nonadiabatic and boundary-value errors, from $O(\epsilon)$ to $O(\epsilon^2)$.
Moreover, the phase shift is mapped onto the mixing angle of the superposition and hence this approach allows one to create arbitrary pre-selected coherent superposition states with very high accuracy.


\subsection{Two pulses}

Let us consider a sequence of two interactions described by two Hamiltonians and two corresponding propagators.
We assume that the first one is parameterized as in Eq.~\eqref{U}, $\mathbf{U}_{1} = \mathbf{U}(t_{\f},t_{\i})$, and the second one $\mathbf{U}_{2}(\phi)$ similarly but with different Cayley-Klein parameters $c$ and $d$ ($|c|^2+|d|^2=1$), and phase-shifted,
\be\label{individual props}
\mathbf{U}_{2}(\phi) = \left[\begin{array}{cc} c& -d^{\ast} e^{-i \phi }  \\ d e^{i \phi } & c^{\ast} \end{array}\right].
\ee
The total propagator then reads
\be\label{total prop}
\mathbf{U} = \mathbf{U}_{2}(\phi) \mathbf{U}_{1}.
\ee
In various cases the second propagator can be related to the first one, and hence expressed by the parameters $a$ and $b$, rather than $c$ and $d$.
These cases include sign flips in the Rabi frequency $\Omega$ or/and the detuning $\Delta$, and also time reversal (i.e. the second pulse is a mirror image of the first one).
In result, the elements of the overall propagator of Eq.~\eqref{total prop} become particularly simple and allow us to create simple recipes for phase control of the quantum dynamics.
Next, we briefly review these relations.


\subsection{Relations between Hamiltonian and propagator changes}

In general, it is not an easy task to relate a change in the Hamiltonian to a change in the propagator.
However, sign flips, phase shifts and time reversal in the former are easily traced in the latter.

(i) \emph{Sign-flip transformations.}
Sign flips of $ \Omega (t)$ and/or $\Delta (t)$ are equivalent to the similarity transformations
\bse\label{H:sigmas relation}
\begin{align}
&\mathbf{H}(t) \stackrel{\Delta \to -\Delta}{\longrightarrow }  \bm\sigma_x\mathbf{H}(t)\bm\sigma_x, \label{H-detuning} \\
&\mathbf{H}(t) \stackrel{\Omega \to -\Omega}{\longrightarrow }  \bm\sigma_z\mathbf{H}(t)\bm\sigma_z, \\
&\mathbf{H}(t) \underset{\Omega \to -\Omega}{\overset{\Delta \to -\Delta}{\longrightarrow }} \bm\sigma_y\mathbf{H}(t)\bm\sigma_y,
\end{align}
\ese
which can easily be derived from Eq.~\eqref{Hamiltonian-Pauli} using the relation $\bm\sigma_k^{-1} = \bm\sigma_k$ $(k=x,y,z)$.

A sign flip transformation of the Hamiltonian, $\mathbf{H}(t)\to \bm\sigma_k\mathbf{H}(t)\bm\sigma_k$, is imprinted onto the propagator as
\be\label{transform:sigmas-general}
\mathbf{U}(t_{\f},t_{\i}) \to \bm\sigma_k\mathbf{U}(t_{\f},t_{\i}) \bm\sigma_k.
\ee
Of special interest is the transformation \eqref{H-detuning}, which flips the sign of the detuning $\Delta$,
\be\label{transform:sigmas-sign flip}
\mathbf{U}(t_{\f},t_{\i}) \stackrel{\Delta \to -\Delta}{\longrightarrow } \bm\sigma_x\mathbf{U}(t_{\f},t_{\i}) \bm\sigma_x = \left[\begin{array}{cc} a^\ast & b \\ -b^{\ast} & a \end{array}\right].
\ee
The sign flip of $\Omega$ is a special case of the phase-jump transformation which follows and therefore we will not consider it separately.

(ii) \emph{Phase jumps.}
A phase jump in the Rabi frequency, $\Omega(t)\to \Omega(t)e^{i \phi }$, can be described with the transformation
\be
\mathbf{H}(t)\to \F(\phi)^\ast \mathbf{H}(t) \F(\phi),
\ee
with
\be
\F(\phi) = e^{i (\phi/2) \sigma_z} = \left[\begin{array}{cc} e^{i \phi/2 } & 0 \\  0 & e^{-i \phi/2 } \\ \end{array}\right].
\ee
The propagator becomes
\be\label{phjump}
\mathbf{U}(t_{\f},t_{\i})\to \F(\phi)^\ast \mathbf{U}(t_{\f},t_{\i})\F(\phi)
=  \left[\begin{array}{cc} a & -b^{\ast} e^{-i\phi} \\  b e^{i\phi} & a^{\ast} \end{array}\right].
\ee
Of course, $\F(\phi)^\ast = \F(-\phi)$.
For $\phi = \pi$ the phase-jump transformation reduces to the $\Omega \to -\Omega$ transformation.

(iii) \emph{Time reflection.} We consider four cases of symmetric and anti-symmetric Rabi frequency and detuning.

1. If $\Omega_2(-t) = \Omega_1(t)$ and $\Delta_2(-t) = \Delta_1(t)$, which means that $\H_2(t) = \H_1(-t)$, then (see Appendix \ref{appendix:time reversal})
\be\label{U2U1-s}
\U_2 = \U_1^T = \left[\begin{array}{cc} a & b \\ -b^\ast & a^\ast \end{array}\right].
\ee

2. If $\Omega_2(-t) = -\Omega_1(t)$ and $\Delta_2(-t) = -\Delta_1(t)$, which means that $\H_2(t) = -\H_1(-t)$, then we have
\be\label{U2U1-a}
\U_2 = \U_1^\dagger = \left[\begin{array}{cc} a^\ast & b^{\ast} \\ -b & a \end{array}\right].
\ee

3. If the Rabi frequency is symmetric and the detuning antisymmetric, i.e., $\Omega_2(-t) = \Omega_1(t)$ and $\Delta_2(-t) = -\Delta_1(t)$, we find from Eq.~\eqref{U2U1-s} or Eq.~\eqref{U2U1-a}, along with Eqs.~\eqref{H:sigmas relation} and \eqref{transform:sigmas-general},  that
\be\label{U2U1-sa}
\U_2 = \bm\sigma_x\U_1^T\bm\sigma_x = \bm\sigma_z\U_1^\dagger\bm\sigma_z = \left[\begin{array}{cc} a^\ast & -b^{\ast} \\ b & a \end{array}\right].
\ee

4. If, on the contrary, $\Omega_2(-t) = -\Omega_1(t)$ and $\Delta_2(-t) = \Delta_1(t)$ we find from Eq.~\eqref{U2U1-s} or Eq.~\eqref{U2U1-a} that
\be\label{U2U1-as}
\U_2 = \bm\sigma_z\U_1^T\bm\sigma_z = \bm\sigma_x\U_1^\dagger\bm\sigma_x = \left[\begin{array}{cc} a & -b \\ b^\ast & a^\ast \end{array}\right].
\ee
Of particular interest are Eqs.~\eqref{U2U1-s} and \eqref{U2U1-sa} because in the other two the sign flip of the Rabi frequency can be included in the phase jump.

\subsection{A pair of pulses with a phase jump}

We consider four cases of sequences of two pulses, as seen in Fig.~\ref{fig:shapes}, with the second one being phase-shifted, i.e. the overall propagator reads $\F(-\phi)\U_2 \F(\phi) \U_1$.
We fix the first pulse to be represented by the propagator $\U_1=\U$ of Eq.~\eqref{U}, and take different choices for the second pulse using the transformations above.
The results are summarized in Table \ref{Table}.
We note here that the first phase gate $\F(-\phi)$ does not affect the transition probability and the simpler sequence $\U_2 \F(\phi) \U_1$ delivers exactly the same transition probability. 


\begin{table}
\begin{tabular}{clclc}
\hline
Pulses & & $\U_2$ & & Transition probability $P$ \\
\hline
\bt{l} $\Omega_2(t) = \Omega_1(t-\tau)$ \\ $\Delta_2(t) = \Delta_1(t-\tau)$ \et
& & $\U_1$
& & $4p(1-p) \cos^2(\alpha+\tfrac12 \phi)$ \\
\hline
\bt{l} $\Omega_2(t) = \Omega_1(t-\tau)$ \\ $\Delta_2(t) = -\Delta_1(t-\tau)$ \et
& & $\sigma_x \U_1 \sigma_x$
& & $4p(1-p) \sin^2(\beta+\tfrac12 \phi)$ \\
\hline
\bt{l} $\Omega_2(t) = \Omega_1(-t)$ \\ $\Delta_2(t) = \Delta_1(-t)$ \et
& & $\U_1^T$
& & $4p(1-p) \sin^2(\alpha-\beta-\tfrac12 \phi)$ \\
\hline
\bt{l} $\Omega_2(t) = \Omega_1(-t)$ \\ $\Delta_2(t) = -\Delta_1(-t)$ \et
& & $\sigma_z \U_1^\dagger \sigma_z$
& & $4p(1-p) \cos^2(\tfrac12 \phi)$
\\ \hline
\end{tabular}
\caption{
Propagator relations for the various cases of pulse pairs illustrated in Fig.~\ref{fig:shapes}.
$\U_1$ and $\U_2$ denote the propagators of the first and second pulses.
When the Rabi frequency $\Omega(t)$ and the detuning $\Delta(t)$ of the second pulse are the same (first row), or are phase shifted (second row), or are mirror images of the ones of the first pulse (third and fourth rows), then the second propagator $\U_2$ can be expressed in terms of the first one $\U_1$ (second column).
The second pulse is phase-shifted, with the resulting propagator $\F(-\phi)\U_2\F(\phi)$.
The overall transition probability for the two-pulse sequence is listed in the third column.
$\alpha$ and $\beta$ are the St\"uckelberg phases of Eq.~\eqref{ab}.
}
\label{Table}
\end{table}

\subsubsection{Identical pulses}

The most natural choice is to take the second pulse the same as the first one, except the phase jump, $\U_2 = \F(-\phi) \U_1 \F(\phi)$, see Fig.~\ref{fig:shapes}(a).
The overall propagator is $\F^\ast \U \F \U$, and the overall transition probability reads
\be\label{p-UfU}
P = 4p(1-p) \cos^2(\alpha + \phi/2).
\ee
Here we have used the polar form of the Cayley-Klein parameters,
\be\label{ab}
a = \sqrt{1-p}\ e^{i \alpha }, \quad b = \sqrt{p}\ e^{i \beta },
\ee
where $p$ is the single-pulse transition probability, while $\alpha$ and $\beta$ are often referred to as the St\"uckelberg phases.
If $p=\frac12$ then $P = \cos^2(\alpha + \phi/2)$, i.e. the transition probability \eqref{p-UfU} depends both on the phase shift $\phi$ and the St\"uckelberg phase $\alpha$.
Therefore, this scenario may not be suitable for robust creation of coherent superposition states unless the phase $\alpha$ is well controlled, which may not be generally the case.

\subsubsection{Bichromatic pulses}

Let us now take the second pulse to have the opposite detuning to the first one, i.e. $\U_2 = \sigma_x \U_1 \sigma_x$, see Fig.~\ref{fig:shapes}(b).
The overall propagator is $\F^\ast \sigma_x \U \sigma_x \F \U$,
and the overall transition probability is given by
\be\label{p-UfU-bichrom}
P = 4p(1-p) \sin^2(\beta + \phi/2).
\ee
If $p=\frac12$ then $P = \sin^2(\beta + \phi/2)$, i.e. the transition probability \eqref{p-UfU-bichrom} depends both on the phase shift $\phi$ and the St\"uckelberg phase $\beta$.
Therefore, as in the preceding case, this scenario may not be suitable for robust creation of coherent superposition states unless the phase $\beta$ is well controlled.

\subsubsection{Time-reflected pulses}

Now take the second pulse to be the mirror image of the first one, i.e. $\Omega(-t) = \Omega(t)$ and $\Delta(-t)=\Delta(t)$, see Fig.~\ref{fig:shapes}(c).
Then $\U_2 = \U_1^T$, see Eq.~\eqref{U2U1-s}.
The overall propagator is $\F^\ast \U^T \F \U$,
and the overall transition probability reads
\be\label{p-UfU-reversed-sym}
P = 4p(1-p) \cos^2(\alpha-\beta-\tfrac12\phi).
\ee
As in the previous two cases, the transition probability \eqref{p-UfU-reversed-sym} depends on the phase jump  $\phi$ and the dynamic phases $\alpha$ and $\beta$.

\subsubsection{Time-reflected bichromatic pulses}\label{sec:reflected_bichrom}

Let us now take the second pulse to be the mirror image of the first one but also the detuning to flip sign, i.e. $\Omega(-t) = \Omega(t)$ and $\Delta(-t)=-\Delta(t)$, see Fig.~\ref{fig:shapes}(d).
Then $\U_2 = \sigma_z \U_1^\dagger \sigma_z$, see Eq.~\eqref{U2U1-sa}.
The overall propagator is $\F^\ast \sigma_z \U^\dagger \sigma_z \F \U$, with Cayley-Klein parameters
\bse
\begin{align}
a_2 &= |a|^2 -|b|^2 e^{-i\phi},\\
b_2 &= ab (1+e^{i\phi}).
\end{align}
\ese
The overall transition probability reads $P=|b_2|^2$, or
\be\label{p-UfU-mirror}
P = 4p(1-p) \cos^2(\tfrac12 \phi).
\ee
Contrary to the previous three cases, the transition probability \eqref{p-UfU-mirror} depends on the phase jump  $\phi$ only, but not on the St\"uckelberg dynamic phases $\alpha$ and $\beta$.
%
Obviously, if $p=\frac12$ then $P = \cos ^2(\phi /2)$.
Therefore, the overall transition probability is determined by the phase jump $\phi$ alone.
In particular, if $\phi = \pi/2$ then $P=\frac12$ and hence a maximally coherent superposition is created, $u^2+v^2 = 1$ and $w=0$.

The Bloch vector components read
\bse\label{Bloch vector}
\begin{align}
u_{\f} &= 4 \sqrt{p(1-p) } \cos(\tfrac12\phi) [(1-p) \cos(\alpha +\beta +\tfrac12 \phi)\notag \\ &-p \cos (\alpha +\beta +\tfrac32 \phi)], \\
v_{\f} &= 4 \sqrt{p (1-p)} \cos (\tfrac12\phi) [ (p-1) \sin (\alpha +\beta +\tfrac12 \phi) \notag \\
 &+ p \sin (\alpha +\beta +\tfrac32 \phi) ], \\
w_{\f} &=  8p(1-p) \cos^2(\phi/2) - 1.
\end{align}
\ese

Such excitation mechanism proves very robust.
Let us assume that each single pulse generates a transition probability close to $\frac12$, i.e. $p = \frac12 - \epsilon$ $(|\epsilon| \ll 1)$.
Then
\be\label{phasepulse:error}
P=4 (\tfrac12 + \epsilon) (\tfrac12 - \epsilon) \cos^2(\phi/2) = (1-4\epsilon^2) \cos ^2(\phi/2).
\ee
Therefore, the error $\epsilon$ in the single-pulse transition probability is relegated to $\mathcal{O}(\epsilon^2)$.
For instance, a deviation of 5\% from the value $\frac12$ in $p$ is reduced to 1\% in the overall transition probability $P$,
 while an error of 1\% in $p$ is reduced to just 0.04\% in $P$.




\subsection{Discussion}

The four cases of pulse pairs split by a phase jump $\phi$ present interesting opportunities.

(i) The fourth case, with the second pulse being a mirror image of the first one and the detuning being an odd function of time, provides a tool for robust creation of arbitrary coherent superpositions of states, with the population ratio controlled solely by the phase $\phi$.
This ratio has no dependence on the dynamical phases of the propagator $\alpha$ and $\beta$.
The robustness derives from the fact that if each constituent pulse creates a transition probability with an error $\epsilon \ll 1$, the sandwiched double pulse gives probabilities with an error $4\epsilon^2$ ($\ll \epsilon$).

(ii) The other three cases produce transition probabilities, which depend on the phase jump $\phi$ as well as on the dynamic phases $\alpha$ [in the first case, Eq.~\eqref{p-UfU}], $\beta$ [in the second case, Eq.~\eqref{p-UfU-bichrom}], or both $\alpha$ and $\beta$ [in the third case, Eq.~\eqref{p-UfU-reversed-sym}].
This renders such configurations inappropriate for controlled creation of superposition states, unless a good control over these phases is possible.
However, we can look at these dependences from another viewpoint: we can use the double-pulse approach to actually \emph{determine} the phases $\alpha$ and $\beta$, associated with a certain interaction.
We note that the error reduction from $O(\epsilon)$ to $O(\epsilon^2)$ occurs in these three cases too because the overall transition probabilities of Eqs.~\eqref{p-UfU}, \eqref{p-UfU-bichrom}, and \eqref{p-UfU-reversed-sym} contain the factor $4p(1-p)$.

We note that the present scenario formally resembles the standard construction of a rotation gate in quantum information in the form of two Hadamard gates split by a phase gate.
However, as the above analysis shows, the relation between the two half-$\pi$ pulses matters.
The first case, when the second pulse is identical to the first one, is the most natural choice; however, if the detuning is nonzero then the overall mixing ratio will depend on the dynamical phase $\alpha$.
(If the detuning is zero then it does not matter if the two half-$\pi$ pulses are identical or not.)
The last case shown in Fig.~\ref{fig:shapes}(d) is the optimal two-pulse arrangement which ensures the maximal accuracy and robustness to errors.


\section{Trigonometric model}\label{sec:three}

In order to illustrate the results in the last section we take the first pulse to be the Cos-Sin model \cite{Zlatanov2017}, in which the Rabi frequency and the detuning are given by
\be\label{cos-sin model}
\Omega(t)=\Omega_0 \cos(t/T), \quad \Delta(t)=-\Delta_0 \sin(t/T),
\ee
with $-\frac12\pi \leqq t/T \leqq 0$.
They are depicted in Figure \ref{fig:shapes} (left frames).
This model has an exact solution if $\Omega_0=\Delta_0=\Lambda$ \cite{Zlatanov2017}, which reads
\bse\label{cos-sin-ab}
\begin{align}
a&=\frac{(1+i A) \sin \left(\frac14\pi s\right) + s \cos \left(\frac14\pi s\right)}{\sqrt{2} s}, \\
b&=\frac{(1-i A) \sin \left(\frac14\pi s\right) -s \cos \left(\frac14\pi s\right) }{\sqrt{2} s},
\end{align}
\ese
where $A=\Lambda T$ is the pulse area and $s=\sqrt{A^2+1}$.
The transition probability reads
\be\label{cos-sin-p}
p = \frac12 - \frac{\sin(\frac12\pi\sqrt{A^2+1})}{2\sqrt{A^2+1}}.
\ee
In the adiabatic limit ($A \gg 1$) we have $p \to \frac12$, with the nonadiabatic oscillations vanishing as $A^{-1}$.

\begin{figure}[tb]
\includegraphics[width=0.90\columnwidth]{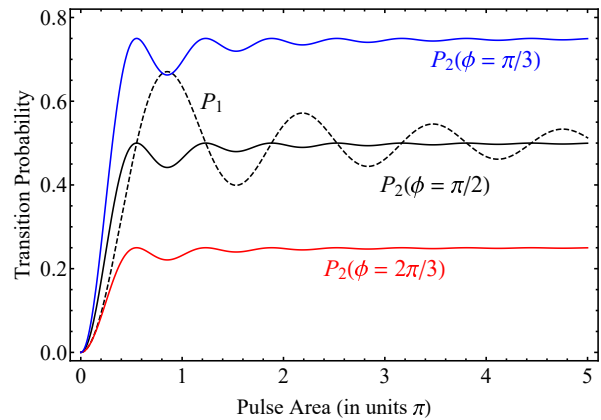}
\caption{(Color online)
Transition probability vs the pulse area $A=\Lambda T$ for the Cos-Sin model of Eq.~\eqref{cos-sin model}
 for a single pulse [dashed, Eq.~\eqref{cos-sin-p}] and for a sequence of two pulses split by phase jumps of $\phi=\frac13\pi$, $\frac12\pi$, and $\frac23\pi$ [solid, Eq.~\eqref{cos-sin-p2}].
In the adiabatic limit ($A \gg 1$), the single-pulse transition probability reaches the value of $\frac12$, while the two-pulse transition probability, depending on the value of the phase jump $\phi$, reaches the values $\frac14$, $\frac12$, and $\frac34$.
}
\label{fig:phi}
\end{figure}

The two-pulse sequence of Fig.~\ref{fig:shapes}(d) with the phase jump of $\phi$ produces 
 the transition probability [cf. Eq.~\eqref{p-UfU-mirror}], which for the Cos-Sin model reads
\be\label{cos-sin-p2}
P = \left[1 - \frac{\sin^2(\frac12\pi\sqrt{A^2+1})}{A^2+1} \right] \cos^2(\tfrac12 \phi).
\ee
In the adiabatic limit ($A \gg 1$) we have
\be\label{cos-sin-p2a}
P \approx \cos^2(\tfrac12 \phi),
\ee
that is the transition probability is determined by the phase shift $\phi$ alone.
In agreement with the general theory [cf. Eq.~\eqref{phasepulse:error}], the nonadiabatic error vanishes as $A^{-2}$, i.e. quadratically faster than for the single pulse, cf. Eqs.~\eqref{cos-sin-p} and \eqref{cos-sin-p2}.
These features are illustrated in Fig.~\ref{fig:phi} where the transition probability for a single pulse (dashed) is compared to double-pulse transition probabilities for phase jumps of $\phi=\frac13\pi$, $\frac12\pi$, and $\frac23\pi$.
Clearly, the nonadiabatic oscillations for the two-pulse sequence are damped much faster.
Moreover, the two-pulse sequence provides the flexibility to reach any desired probability ($\frac14$, $\frac12$, and $\frac34$ in this figure) in the adiabatic limit.

\section{Extension to longer sequences}\label{sec:multi}

\begin{figure}[tb]
\bt{c}
\bt{cc} \includegraphics[width=0.25\columnwidth]{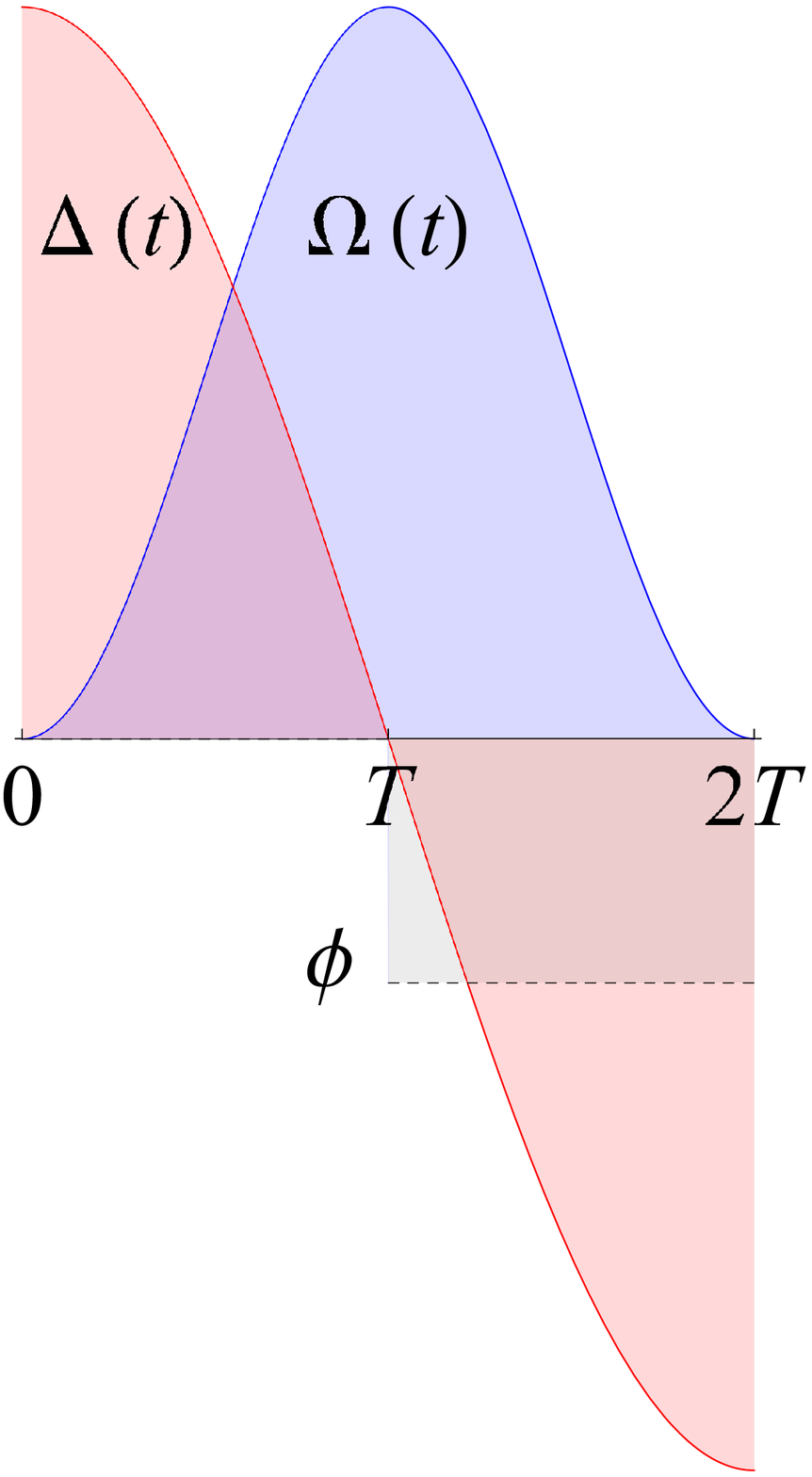} & \hspace{10mm} \includegraphics[width=0.50\columnwidth]{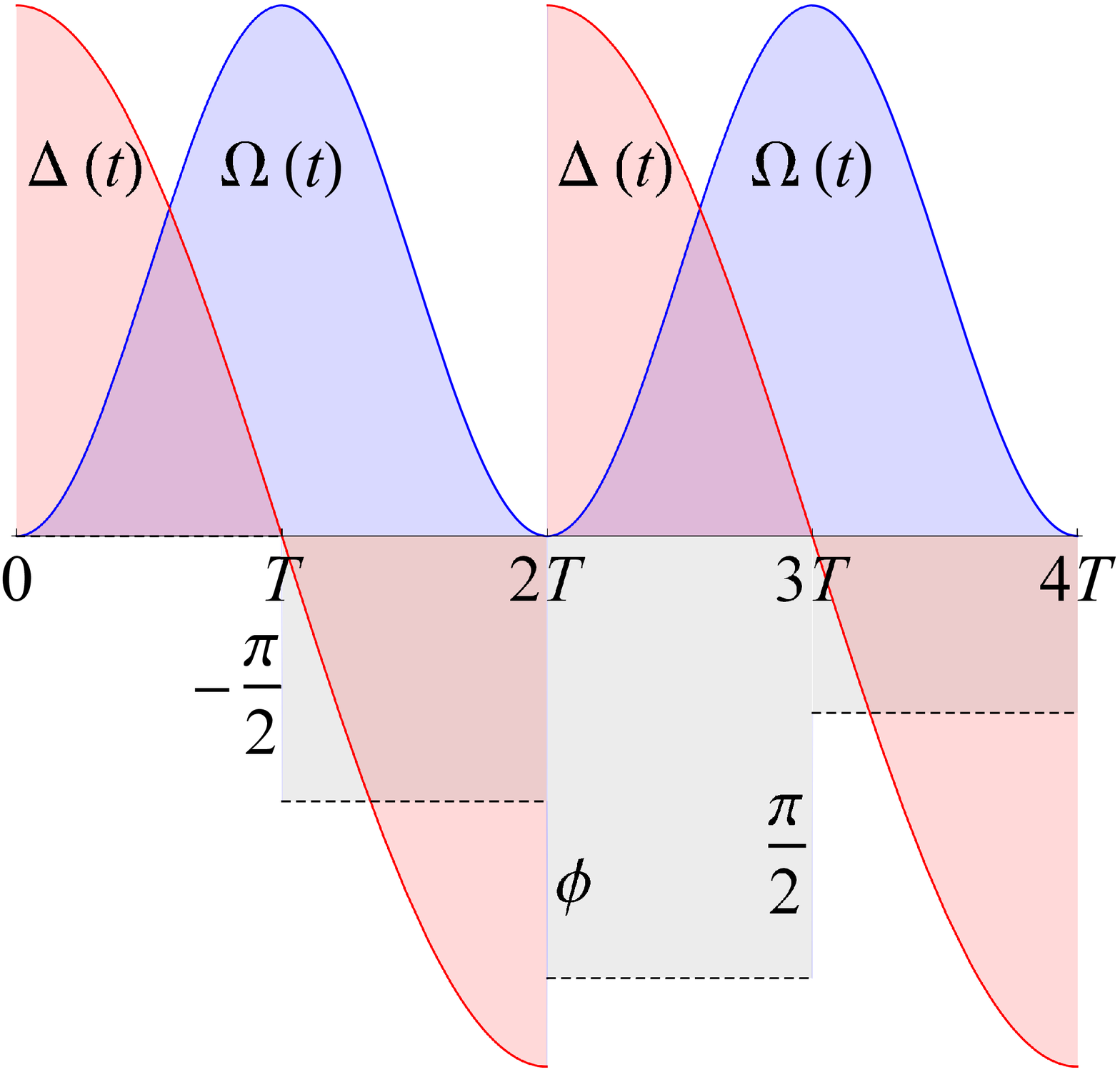} \et \\
\includegraphics[width=1.00\columnwidth]{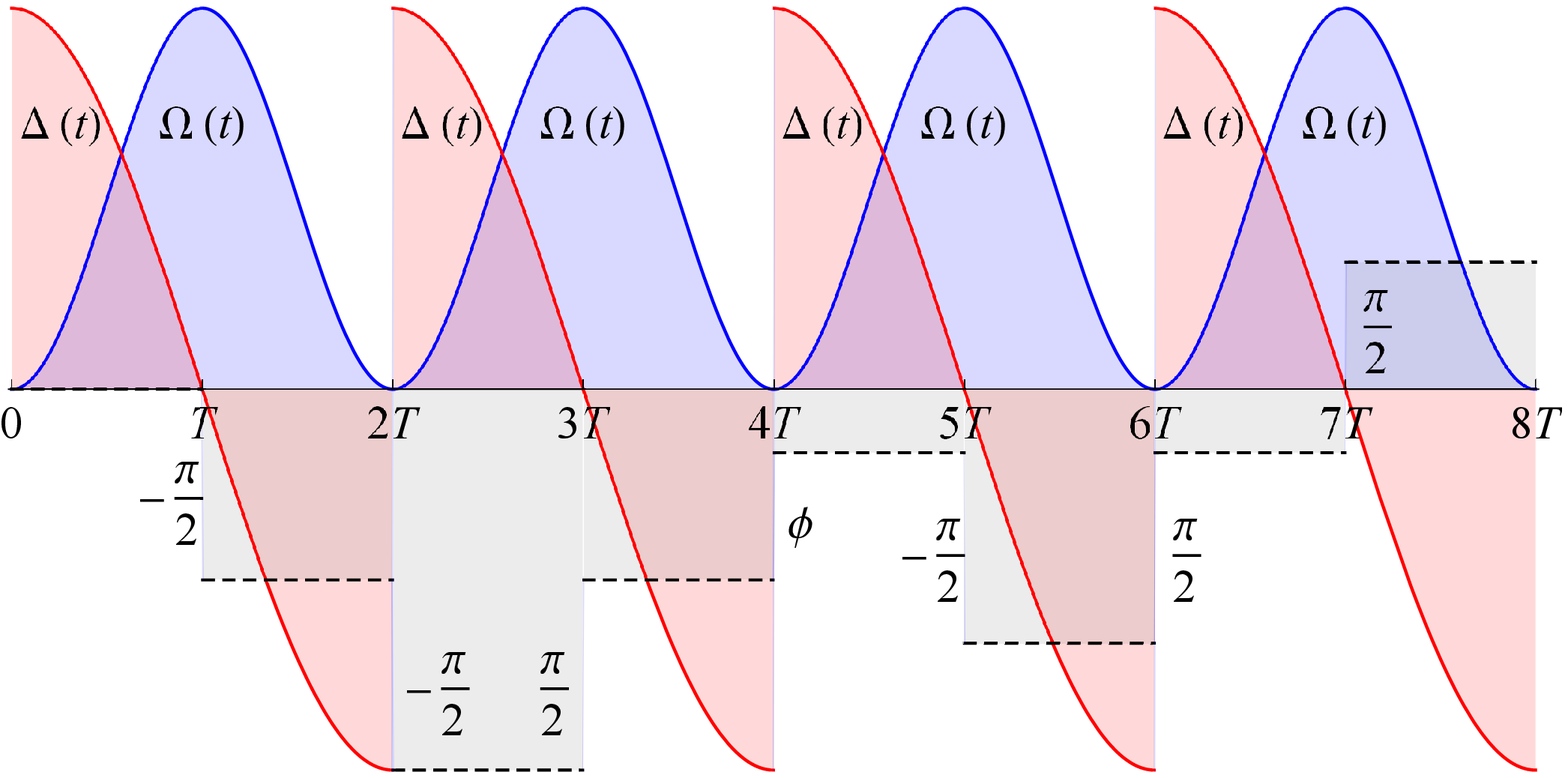}
\et
\caption{(Color online)
Concatenated sequences of $N=2$, 4 and 8 pulses, which produce a transition probability of $\cos^2(\phi/2)$ with an error $O(\epsilon^N)$.
}\label{fig:multi}
\end{figure}

The two-pulse technique can be extended to longer sequences thereby further reducing the error.
We use an iterative nesting technique as follows.
As discussed hitherto, a pair of two pulses, which produce a coherent superposition state with a mixing angle $\phi$, reads
\be\label{U2}
\U_2(\phi) = \overline{\U} \F(\phi) \U,
\ee
where $\overline{\U} = \sigma_z \U^\dagger \sigma_z$, as before (Sec.~\ref{sec:reflected_bichrom}).
For a target transition probability of $\frac12$ we set $\phi=\frac12\pi$.
We use this sandwich in order to construct the next sequence by replacing each of the pulses $\U$ by the sandwich $\U_2(\frac12\pi)$:
\be\label{U4}
\U_4(\phi) = \overline{\U} \bm{\Pi}^* \U \F(\phi) \overline{\U} \bm{\Pi} \U,
\ee
where $\bm{\Pi} = \F(\frac12\pi)$ and $\bm{\Pi}^* = \F(-\frac12\pi)$.
We continue by replacing $\U$ by $\U_4(\pi/2)$ in Eq.~\eqref{U2} to find
\be\label{U8}
\U_8(\phi) =
\overline{\U} \bm{\Pi}^* \U \bm{\Pi}^* \overline{\U} \bm{\Pi} \U
\F(\phi)
\overline{\U} \bm{\Pi}^* \U \bm{\Pi} \overline{\U} \bm{\Pi} \U.
\ee
Then we replace $\U$ by $\U_8(\pi/2)$ in Eq.~\eqref{U2} to obtain the next sequence,
\begin{align}\label{U16}
\U_{16}(\phi) &=
\overline{\U} \bm{\Pi} \U \bm{\Pi} \overline{\U} \bm{\Pi}^* \U \bm{\Pi}^* \overline{\U} \bm{\Pi} \U \bm{\Pi}^* \overline{\U} \bm{\Pi}^* \U
\F(\phi) \notag \\
&\times \overline{\U} \bm{\Pi}^* \U \bm{\Pi}^* \overline{\U} \bm{\Pi} \U \bm{\Pi} \overline{\U} \bm{\Pi}^* \U \bm{\Pi} \overline{\U} \bm{\Pi} \U,
\end{align}
and so on.
This concatenation procedure produces sequences of $N=2^n$ pulses.
The first few concatenated sequences are shown in Fig.~\ref{fig:multi}.

Consider a target transition probability of $\frac12$.
The concatenation procedure described above leads to the following transition probability
\be
P_N = \frac12 \left[1 - (1-2p)^N \right].
\ee
If $p=\frac12-\epsilon$ then $P_N = \frac12 \left[1 - (2\epsilon)^N \right]$.
Therefore, the relative error scales as $(2\epsilon)^N$.
In other words, if a single pulse produces a transition probability 0.45 (instead of $\frac12$, meaning $\epsilon=0.05$ or 10\% error) then the sequence composed of two such pulses
will reduce the probability error to 1\%, the four-pulse sequence will further reduce the error to 0.01\%, and the eight-pulse sequence to $10^{-6}$.
Instead, if the single pulse produces a transition probability of 0.49 (meaning $\epsilon=0.01$ or 2\% error), then the two-pulse sequence will reduce the error to 0.04\%, the four-pulse sequence to $1.6 \times 10^{-7}$, etc.
Hence the concatenated-sequence procedure allows one to quickly reduce the probability error beyond the quantum computation benchmark values (usually $10^{-4}$), even when starting with low-fidelity pulses.

\begin{figure}[tb]
\bt{c}
\includegraphics[width=0.85\columnwidth]{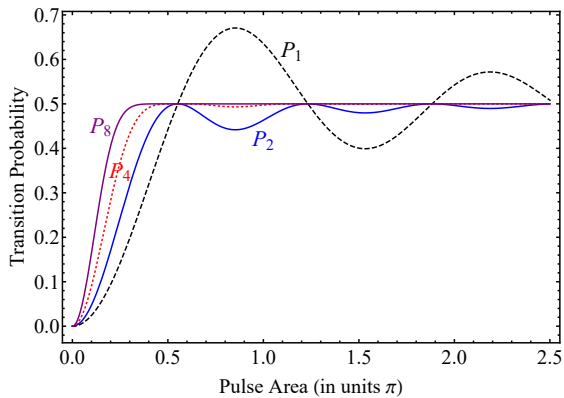}
\et
\caption{(Color online)
Transition probabilities $P_N$ of Eq.~\eqref{PN-cos-sin} for the multi-pulse sequences of Fig.~\ref{fig:multi} compared to the single-pulse transition probability $P_1$.
}\label{fig:multi-p}
\end{figure}

For the sin-cos model the transition probability for the concatenated sequences $\U_{N}(\phi)$ reads
\be\label{PN-cos-sin}
P_N = \frac12 \left[ 1 - \frac{\sin^N(\frac12\pi\sqrt{A^2+1})}{(A^2+1)^{N/2}} \right]
\ee
The probability error diminishes very quickly with $N$: as $A^{-N}$.
Figure \ref{fig:multi-p}, which shows the transition probability versus the pulse area $A$ for a single pulse and sequences of 2, 4 and 8 pulses, demonstrates this scaling.
The nonadiabatic oscillations for a single pulse require very large values of the pulse area in order to be diminished to sufficiently small values because their amplitude scales as $A^{-1}$.
The oscillation damping is much faster already for two pulses (as $A^{-2}$), while for sequences of four (as $A^{-4}$) and eight (as $A^{-8}$) pulses oscillations are barely visible at all.
We stress that this error damping does not derive merely from the larger total pulse area of the sequences but rather from the effect of destructive cancellation of errors, similar to what happens in composite pulses.
For example, the four-pulse sequence has obviously a factor of 4 larger pulse area than a single pulse.
However, the error damping produced by the four-pulse sequence for $A>0.4\pi$ is far stronger than the error damping by a single pulse for $A>4\times 0.4\pi = 1.6\pi$.

\section{Summary, conclusions and outlook}\label{sec:conclusions}

In this paper we proposed a technique for accurate, flexible and robust generation of coherent superposition qubit states.
The technique uses a pair of adiabatic pulses split by a phase jump, which is used as a control parameter.
Each pulse uses a chirped detuning which induces a half crossing that acts approximately as a half-$\pi$ pulse in the adiabatic regime.
The phase jump is directly mapped onto the mixing angle of the created superposition state.
The error $\epsilon \ll 1$ of each half-$\pi$ pulse is suppressed to $4\epsilon^2 (\ll \epsilon)$ by the two-pulse sequence, thereby allowing to improve the overall accuracy.
In such a manner we combine the benefits of \emph{robustness} stemming from adiabatic evolution with \emph{accuracy} generated by the error suppression, and \emph{flexibility} of the created superposition state the population ratio of which is determined by the value of the phase jump $\phi$.
In particular, a maximally coherent superpositions with equal populations is created for $\phi=\frac12\pi$.

Given the first pulse, we have identified four distinctly different choices for the second pulse, as depicted in Fig.~\ref{fig:shapes}.
The Rabi frequency of the second pulse is the same as the first pulse or a mirror image of it, and the same applies to the detuning, with the latter having the same or the opposite sign to the first pulse.
We have shown that in the general case, only one of these four cases can be uses for the proposed technique, namely, when the Rabi frequency and the detuning of the second pulse are mirror images of those of the first pulse, and the two detunings have opposite signs, see Fig.~\ref{fig:shapes}(d).
Then the overall transition probability depends on the phase jump $\phi$ only.
In the other three cases, it depends also on the dynamic phases of the propagators.
Therefore, the sequence of Fig.~\ref{fig:shapes}(d) can be used for efficient, flexible and robust creation of pre-selected superposition states, while the other three sequences in Fig.~\ref{fig:shapes} can be used for tomography of coherent superpositions.

The proposed technique formally resembles the well known sequence of two Hadamard gates split by a phase gate for creating an arbitrary rotation gate.
The present analysis shows that in the general case of asymmetric temporal shape of the Rabi frequency and nonzero detuning, the most obvious scenario of using two identical pulses, as in Fig.~\ref{fig:shapes}(a), is not the optimal one because then the rotation angle is shifted by a (probably unknown) dynamical phase.
It is only the last case (d) in Fig.~\ref{fig:shapes}, which eliminates such unwanted shifts.

We used a simple exactly soluble trigonometric model to illustrate the proposed quantum control technique.
It allows one to explicitly estimate the nonadiabatic oscillations and their damping in the near-adiabatic regime.
This damping behaves as $A^{-1}$ for a single pulse and $A^{-2}$ for the two-pulse sequence.
The analytic model shows that a high-fidelity rotation gate can be generated by sequences of pulses with areas of just over $\pi$.

The proposed technique was extended to sequences of more than two pulses by concatenating half-$\pi$ sequences and splitting them by a phase jump.
The two concatenated sequences must obey the general symmetry principles of the two-pulse sequence:
 the Rabi frequency and the detuning of the second sequence should be the mirror images relative to the ones of the first sequence, and the detuning must also flip its sign.
In this manner, we obtain sequences of $N=2^n$ pulses, with the nonadiabatic error $\epsilon$ scaling as $(2\epsilon)^{N}$.
Therefore, starting from a low-fidelity pulse with a significant error $\epsilon$, one can achieve very high error correction by appropriately concatenating this pulse.
This makes the proposed technique appropriate for generating very high-fidelity quantum rotation gates, such as the Hadamard gate, with rather poor initial resources.
Moreover, this approach is applicable to a wide range of systems, including ground-state atomic \cite{atoms} and ionic \cite{ions} qubits, Rydberg atoms \cite{Saffman2010}, Rydberg ions \cite{Higgins2017}, superconducting qubits \cite{Devoret2013}, etc.

Finally, we have focused at pulse pairs obtained by twinning two half-crossing adiabatic pulses.
The same approach can be used if each $\pi/2$ is produced in a different manner, e.g. by resonant or detuned fields.
The accuracy, the robustness and the flexibility of the resulting sequence may be similar to the present work provided the symmetry relations of Fig.~\ref{fig:shapes} are satisfied.
This makes the present approach applicable beyond the adiabatic regime.



\acknowledgments
The authors thank Marcel Hein for his experimental insights on the paper.
This work has been partly supported by the European Unions's Horizon 2020 research and innovation programme under the Marie-Curie grant agreement No. 641272 and by the Bulgarian Science Fund Grant DO02/3 (ERyQSenS).

\appendix

\section{Time reversal}\label{appendix:time reversal}

If we change the direction of time to $t\to-t$ we see that Eq.~\eqref{SEq-U} changes to
\be\label{SEq-t}
{\rm i \hbar}\frac{d}{dt}\U(-t,0) = -\mathbf{H}(-t)\U(-t,0).
\ee
There are important differences between symmetric and anti-symmetric Hamiltonians.

(i) The case of symmetric Hamiltonian $\H(t)=\H(-t)$ is obtained if $\Omega(t)=\Omega(-t)$ and $\Delta(t)=\Delta(-t)$.
If we complex conjugate Eq.~\eqref{SEq-U} we find
\be\label{CC schrod}
{\rm i \hbar}\frac{d}{dt}\U(t,0)^\ast = -\mathbf{H}(t)\U(t,0)^\ast.
\ee
provided $\mathbf{H}(t)$ is real.
The initial condition for both $\U(t,0)^\ast$ and $\U(-t,0)$ is the same: $\U(0,0)^\ast = \U(0,0) = \I.$
Then $\U(t,0)^\ast$ and $\U(-t,0)$ coincide because they satisfy the same differential equation with the same initial condition,
\be
\U(-t,0)=\U(t,0)^\ast = \left[\begin{array}{cc} a^{\ast} & -b \\ b^{\ast} & a \end{array}\right].
\ee

(ii) The anti-symmetric case $\H(-t)=-\H(t)$ implies $\Omega(-t)=-\Omega(t)$ and $\Delta(-t)=-\Delta(t)$.
Then Eq.~\eqref{SEq-t} will have the same form as Eq.~\eqref{SEq-U} and therefore $\U(-t,0)$ and $\U(t,0)$ will be equal 
\be
\U(-t,0) = \U(t,0) =  \left[\begin{array}{cc} a & -b^{\ast} \\ b & a^{\ast} \end{array}\right].
\ee

In both symmetric and anti-symmetric cases we have due to unitarity
\be
\U(0,-t) = \U(-t,0)^\dagger.
\ee
Therefore, for a symmetric $\H(t)$ we have
\be
\U(0,-t) = \U(t,0)^T = \left[\begin{array}{cc} a & b \\ -b^\ast & a^\ast \end{array}\right],
\ee
while for antisymmetric $\H(t)$ we find
\be
\U(0,-t) = \U(t,0)^\dagger = \left[\begin{array}{cc} a^\ast & b^{\ast} \\ -b & a \end{array}\right].
\ee

If $\Omega(-t) = \Omega(t)$ and $\Delta(-t) = -\Delta(t)$  we have
\be
\U(0,-t) = \sigma_z\U(t,0)^\dagger\sigma_z = \left[\begin{array}{cc} a^\ast & -b^{\ast} \\ b & a \end{array}\right].
\ee

If $\Omega(-t) = -\Omega(t)$ and $\Delta(-t) = \Delta(t)$  we have
\be
\U(0,-t) = \sigma_z\U(t,0)^T\sigma_z = \left[\begin{array}{cc} a & -b \\ b^\ast & a^\ast \end{array}\right],
\ee






\Black

\end{document}